\documentclass[12pt,reqno]{amsart}
\newtheorem{thm}{Theorem}[section]
\newtheorem{prop}[thm]{Proposition}

\newtheorem{defn}[thm]{Definition}
\makeatletter \@addtoreset{equation}{section} \makeatother

\newcommand{\W}{\mbox{${\mathfrak W}$}}
\newcommand{\wat}[1]{\mbox{$\mathfrak{W}(\mathfrak{A}^{#1})$}}

\newcommand{\winf}{\mbox{${\mathfrak W}^{\infty}$}}

\newcommand{\threads}{\mbox{$E_{\infty}$}}

\newcommand{\C}{\mbox{${\mathcal C}(X)$}}

\newcommand{\glosta}{\mbox{$\mathcal{K}\C$}}
\newcommand{\R}{\mbox{$\mathbf{\sf R}$}}
\newcommand{\zemu}{\mbox{$\zeta_{\mu}$}}
\newcommand{\B}{\mbox{$\mathcal{B}$}}
\newcommand{\Bo}{\mbox{${\mathcal B}(X)$}}

\newcommand{\chip}[1]{\mbox{$\chi^{(X)}_{#1}$}}
\newcommand{\Q}{\mbox{${\mathfrak P}$}}

\newcommand{\spg}{\vspace{5mm} \noindent}
\newcommand{\hk}{\mbox{Haag-Kastler }}
\newcommand{\prf}{\mbox{{\em Proof.\hspace{2mm}}}}
\newcommand{\qd}{\mbox{\hspace{5mm}\rule{2.4mm}{2.4mm}}}

\begin{document}
\begin{center}
{\huge {\bf An algebraic theory of infinite classical lattices II:

\vspace{4mm}Axiomatic theory}}

\vspace{20mm} {\bf Don Ridgeway}

\vspace{15mm}
Department of Statistics,\\
North Carolina State University,\\
Raleigh, NC  27695\\
ridgeway@stat.ncsu.edu
\end{center}

\vspace{1cm} \hspace{53mm} {\bf Abstract}

{\small  We apply the algebraic theory of infinite classical lattices from Part I to
write an axiomatic theory of measurements, based on Mackey's axioms for quantum
mechanics.  The axioms give a complete theory of measurements in the sense of Haag and
Kastler, taking the traditional form of a logic of propositions provided with a classical
spectral theorem.  The results are expressed in terms of probability distributions of
individual measurements. As applications, we give a separation theorem for states by the
set of observables and discuss its relationship to the equivalence of ensembles in the
thermodynamic-limit program. We also introduce a weak equivalence of states based on the
theory.}

\vspace{1cm} \noindent MSC 46A13 (primary), 46M40 (secondary)

\newpage
\section{Introduction.}

\vspace{5mm} \noindent{\bf {\large I  \hspace{4mm} Introduction}} \setcounter{section}{1}
\setcounter{equation}{0} \setcounter{thm}{0}

There are two standard   approaches to the study of infinite lattice systems,  the
algebraic approach from quantum field theory (QFT) (\cite{brat87}, \cite{emch72},
\cite{emch84}, \cite{sega47} ) and  the thermodynamic limit (TL) (\cite{lanf73},
\cite{ruel69} ).  In Part I of this series \cite{ridg05}, we presented an algebraic
theory of infinite classical lattices, constructed using the axioms of Haag and Kastler
\cite{haag64} from the QFT. We showed that the two approaches may be regarded as two
aspects of a single theory, linked by a unique relation between their states based on
expectation values. The kinds of questions they can ask are different, however. TL theory
is designed to study states,  especially the equilibrium states, and the expectation
values they assign to observables. We shall find that with the algebraic theory, one may
study the statistical properties of the individual measurement. This is therefore the
setting for a theory of measurement.

In this paper, we show that the abstract system of algebraic observables of the theory
satisfies the Mackey axioms I-VI from quantum mechanics. This will permit us to base the
axiomatization on a logic of propositions provided with a classical spectral theorem.
Following Birkhoff and von Neumann, the theory is then centered on the question, ``If I
measure a certain quantity on a lattice prepared in a given state, what is the
probability the outcome will lie in a fixed  interval $(a,b)$?'' (\cite{birk36},
\cite{jauc69}, \cite{mack63}).

\vspace{5mm} \noindent{\bf {\large II \hspace{4mm} Classical measurements }}
\setcounter{section}{2} \setcounter{equation}{0} \setcounter{thm}{0}

\begin{center} {\bf A. The Haag-Kastler frame.} \end{center}

In agreement with Haag and Kastler, we should treat measurements such that their ``state
and operation are defined in terms of laboratory procedures'' \cite[p.850]{haag64}. For
this purpose, we view the lattice as representing a finite system immersed homogeneously
in an (infinite) surround which acts as a generalized temperature bath. We denote by
$\mathcal P$ the set of all possible systems, indexed by $\mathbf J$.

According to the \hk axioms, the algebraic structure is derived from the local {\em
texture}, {\em i.e.,} the pairing of each system with the set of functions representing
measurements on that system.   We denote the configuration space of the lattice by
$\Omega$, written as the Cartesian product of the single-site configurations, so that for
any system $\Lambda_t$, $t \in \mathbf J$, we may write $\Omega = \Omega_{\Lambda_t}
\times \Omega_{\Lambda_t^{\prime}}$. To each system $\Lambda_t$, we assign the set
$\wat{t}$ of   functions on $\Omega$ representing measurements on  $\Lambda_t$  and the
compact set $E_t$ of states on $\wat{t}$.

The axioms then direct formation of the algebraic theory from the texture.  We showed
that for any compact convex set $K$ of algebraic states, we may construct the triple
$\{X, \C, \glosta \}$, dependent on $K$, where the Segal algebra $\C$ is the set of
continuous functions on a compact space $X$, and $\glosta$ is the set of states on $\C$.
The triple has the following structure:

\begin{enumerate}

\item $\glosta$ is isomorphic with $K$.

\item All of the local functions $(\wat{t})$ representing measurements on finite systems
of the lattice map to unique points in $\C$.  Functions measuring the same physical
quantity on different systems map to the same point in $\C$.

\item  $X$ is homeomorphic with the set $\partial_e \glosta$ of extremal points of
$\glosta$.

\item By the Riesz representation theorem, for every state $\zeta \in \glosta$, there
exists a unique Radon probability measure $\sigma$ on $X$ such that

\begin{equation} \zeta(f)  = \int_X f(x) d\sigma(x) \hspace{3mm} \forall f \in \C
\end{equation}

\spg This is formally the expectation value of a point $f \in \C$ for a lattice in state
$\zeta$. Physically, it is the expectation value of any local measurement that maps to
$f$. Note especially  that this is a decomposition theorem, {\em i.e.} it decomposes any
state $\zeta \in \glosta$ into an integral over the pure states of $\glosta$.

\end{enumerate}

\noindent In this paper we shall only be concerned with a particular choice of $K$, the
compact convex set of all stationary states of the lattice. For this case, we showed the
crucial additional fact that

\vspace{3mm} \hspace{1mm} (5) the space $X$ is a  Stonean  (compact extremely
disconnected) topolog- \vspace{3mm} \hspace{11mm} ical space (Theorem V.3).

The point is that this triple is an algebraic theory well-defined by these five
properties {\em without any reference to an underlying structure.}  We apply Mackey's
axioms to {\em this} structure and derive our theory of measurements in terms of  it.

The similarity of eq. (2.1) to the integration over phase space in ordinary CSM to obtain
expectation values might lead to the question of whether the Mackey axioms could be
applied directly to the classical problem with its configurational phase space. However,
usually the set of continuous functions is  not large enough to represent the observables
of a classical problem.  For example, the $(\wat{t})$ are the sets of all bounded
measurable functions compatible with the preparation of systems for measurement.
Furthermore,  a phase space with a Stonean topology, which will be essential in the
following, excludes most interesting mechanical problems.

We shall have occasion to use a term ``microcanonical state'' for the lattice. This term
clearly refers to local states. There are two ways of describing states of the infinite
system. The one is in terms of $\glosta$, the states (positive linear functionals of norm
1) on the algebra $\C$. Part I gives another way, namely, in terms of an  inverse limit
object of the $((E_t)_{t \in J})$, denoted there by $\threads$. The elements of
$\threads$ are threads $(\mu_t)_{t \in J}$ giving the local state of each finite system
in the lattice. It is shown in Part I that the two sets $\glosta$ and $\threads$ are
isomorphic. The ``microcanonical state'' refers to a state on $\C$ identified with a
thread $(\mu_t) \in \threads$ in which all local states  are microcanonical.

\vspace{3mm} \begin{center} {\bf B. Measurements} \end{center}

We adopt Segal's  interpretation of the algebraic observables.   In Segal's terminology,
the elements of $\C$ are the {\em observables}, and the values $f(x),$ $ x \in X,$ the
{\em spectral values} of $f \in \C$. They are the only possible values of any measurement
$f^t \in \mathfrak{W}(\mathfrak{A}^t)$ representable by $f$.   The mathematical states
$\zemu \in \glosta$ define ensembles or {\em distributions} of the {\em pure states} $X$,
so that the {\em expectation values} of measurements are the quantities $\zemu(f)$.

A description based on this terminology requires something conceptually close to the
following algebraic picture of the classical measurement.   In the preparation for
measurement, the lattice is brought into a given state $\zeta \in \glosta$. The
measurement begins with an instantaneous isolation that leaves it in a MC state $x_{\mu}
\in X $ randomly chosen from the ensemble defined by $\zeta$.   The {\em outcome} of the
measurement $f \in \C$ is the MC average $f(x_{\mu})$, the result of time averaging, say.
Its {\em expectation value} is $\zeta(f)$, the integral over the possible outcomes $x \in
X$.

\vspace{5mm} \noindent{\bf {\large III \hspace{4mm} Mackey's axioms  }}
\setcounter{section}{3} \setcounter{equation}{0} \setcounter{thm}{0}

This description of a measurement is readily turned into an axiomatic theory based on
Mackey's axioms for a quantum theory \cite{mack63}. It will have  the traditional form of a
logic of propositions introduced by Birkhoff and von Neumann (\cite{birk36}). For the
commutative case, a Mackey system is defined by six axioms.   In the following, we
construct such a system from our triple $\{X, \C, \mathcal{K}\C\}$.

A theorem in Mackey gives sufficient conditions for a lattice of observables and its states
to display his six axioms \cite[p.68]{mack63}. The next two propositions satisfy these
conditions. The first pertains to observables.

\begin{prop} The mapping $\phi: \Q \rightarrow \B(X)$ by $\phi(\chip{F}) = F$ is a
lattice-isomorphism from the class $\Q$ of idempotents of $\C$ onto the topology ${\mathcal
B}(X)$ of $X$.  Hence, $\Q$ is a complete Boolean algebra.
\end{prop}

\noindent \prf Observe first that $\chip{F} \in \C$ iff $F$ is clopen (=closed-and-open).
It was shown in Part I that all open sets are clopen (Theorem VI.3). Hence, the
idempotents are exactly the characteristic functions $\chip{F}$, $F \in {\mathcal B}(X)$.
Then $\phi$ is clearly 1:1 and onto, i.e., $\Q = {\mathcal B}(X)$. Also, $\forall E, F
\in \B(X)$, $\phi(\chip{E} \bigvee \chip{F}) = \phi(\chip{E \bigcup F}) = E \bigcup F =
\phi(\chip{E}) \bigcup \phi(\chip{F})$, and $\phi(\chip{E} \bigwedge \chip{F}) =
\phi(\chip{E \bigcap F}) = E \bigcap F = \phi(\chip{E})$ $ \bigcap \phi(\chip{F})$. The
complementation is defined by $(\chip{F})^{\prime} = {\bf 1} - \chip{F} =
\chip{F^{\prime}}$ and hence $\phi((\chip{F})^{\prime}) = F^{\prime}$.  Hence, $\phi$ is
a lattice isomorphism.  For completeness, note simply that for an arbitrary net
$(\chip{F_i})$, $\bigcup F_i \in {\mathcal B}(X)$ (clopen), so that $\bigvee \chip{F_i} =
\chip{\bigcup F_i} \in \Q$. One shows similarly that the lattice $\Q$ is distributive.
The distributive property for infinite operations is given by Semadeni \cite[Proposition
16.6.3]{sema71} \hspace{5mm} \qd

The completeness of the lattice $\Q$ is equivalent to having $X$  Stonean
\cite[6.2.4]{port88}, as pointed out in Part I.  It is analogous to the completeness of
the lattice of projections of the von Neumann algebra in algebraic QFT \cite[Proposition
V.1.1]{take79}. The distributive lattices are exactly those with a set representation
\cite[Birkhoff-Stone theorem, p. 104]{port88}. Since the distributive property assures
the pairwise compatibility of measurements, this Proposition  assures that we are dealing
with classical theory.

The second condition pertains to states.

\begin{prop}  Denote by $\mathcal{S}$ the set of all restrictions
$\{ \zemu|_{\Q}, \mu \in \threads\}$. Then $\mathcal{S}$ is a full and strongly convex set
of states on $\Q$.
\end{prop}

\spg {\em Proof.} The state $\zemu \in \mathcal{S}$ is a state on $\Q$ in Mackey's sense
if, in addition to $\zemu (\chi^{(X)}_{\emptyset}) = 0$ and $\zemu(\chi^{(X)}_X) = 1$, one
has that for all sets of questions $(\chi^{(X)}_{F_n} ) \in \Q$ with $F_i \bigcap F_j =
\emptyset \hspace{2mm} \forall i \neq j$, $\zemu(\bigvee \chi^{(X)}_{F_n} ) =
\zemu(\chip{\bigcup F_n}) = \sum \zemu (\chi^{(X)}_{F_n})$. Certainly for all finite
subsets of $(\chi^ {(X)}_{F_n} )$, $\zemu(\bigvee_{i=1}^k \chi^{(X)}_{F_{n_i}} ) =
\sum_{i=1}^k \zemu (\chi^{(X)}_{F_{n_i}})$. The result then follows by continuity. Now note
that for any pair $\chi^{(X)}_E$, $\chi^{(X)}_F$, if $\chi^{(X)}_E$ is not $ \leq
\chi^{(X)}_F$, then there exists $x_{\mu} \in X$ such that $\chi^{(X)}_E(x_{\mu}) = 1$ and
$\chi^{(X)}_F(x_{\mu}) = 0.$ But $\delta_{x_{\mu}} \in \glosta$, while
$\delta_{x_{\mu}}(\chi^{(X)}_E) = 1$ and $\delta_{x_{\mu}}(\chi^{(X)}_F) = 0$. Hence,
$\mathcal{S}$ is a {\em full} set of states, i.e., if $\zemu(\chi^{(X)}_E) \leq
\zemu(\chi^{(X)}_F)$ for all $\zemu \in \mathcal{S}$, then $\chi^{(X)}_E \leq
\chi^{(X)}_F$. Finally, the set of states $\mathcal{S}$ is {\em strongly convex} in
Mackey's sense if for any sequence $(t_n) \in [0,1]$ such that $\sum _1^{\infty} t_n = 1$
and any set $(\zeta_{\mu_n}) \in \mathcal{S}$, $\sum_1^{\infty} t_n \zeta_{\mu_n} \in
\mathcal{S}$. Certainly $\sum_1^{\infty} t_n \zeta_{\mu_n}$ is a positive linear functional
on $\C$ by continuity. Furthermore, $\| \sum_1^{\infty} t_n \zeta_{\mu_n} \| = \sup_{ \| f
\| \leq 1} \sum_1^{\infty} t_n \zeta_{\mu_n}(f) = \sum_1^{\infty} t_n \zeta_{\mu_n}(\chi_X)
=  \sum_1^{\infty} t_n = 1.$ Hence, $\sum_1^{\infty} t_n \zeta_{\mu_n} \in \glosta$.   \qd

\vspace{3mm} The Mackey axioms are in terms of a class of functions of the following form.

\begin{defn} Denote by $\B$ the Borel sets of the real line $\R$.
The function $Q:{\mathcal B} \rightarrow \Q$ is called a {\bf {\em $\Q$-valued measure}} on
$\R$ iff the following obtain:

\noindent (a) $ Q(\emptyset) = 0,$ $ Q(${\bf {\sf R}}) = 1;

\noindent (b) If $(B_n)$ is any family in ${\mathcal B}$, and $B_i \cap B_j = \emptyset
\hspace{2mm}\forall i \neq j$, then $Q(\cup B_n) = \bigvee Q(B_n)$.
\end{defn}

\spg Note that $\bigvee Q(B_n) \in \Q$ because $\Q$ is complete.

Let $\mathcal O$ be the set of all $\Q$-valued measures on $(\R, \B)$. $\mathcal O$ is the
set of observables of the  Mackey system $(\mathcal{O}, \mathcal{S}, \B)$. With
Propositions III.1 and III.2, we have proven the following.

\begin{thm}   The triple $\{\mathcal{O},
\mathcal{S}, \B \}$ is a Mackey system, satisfying  Axioms I - VI.  \qd
\end{thm}

\spg It is noteworthy that with $\Q$ a complete lattice, the system  $\{X, \C,
\mathcal{K}\C\}$ likewise satisfies the axioms of Piron from QFT \cite{piro64}.

\vspace{5mm} \noindent{\bf {\large IV \hspace{4mm} The theory of measurement  }}
\setcounter{section}{4} \setcounter{equation}{0} \setcounter{thm}{0}

We divide discussion into two sections, dealing respectively with observables
and states.

\vspace{3mm} \begin{center} {\bf A. Observables} \end{center}

The role of the quasilocal observables depends on their identification with the elements
of $\mathcal O$. Observe first that $\mathcal{O}$ is a large set. In fact, if $f \in \C$
is any observable, and $B \in {\mathcal B}$, define $Q^f: \B \rightarrow \mathbf{Q}$ by
$Q^f(B) \equiv Q^f_B = \chi_B \circ f = \chip{[f \in B]}$. Recall that $\bigvee
\chip{B_i} = \chip{\cup B_i}$. Hence $Q^f \in \mathcal{O}$. Axiom VI says that all of
$\mathcal O$ is of this form:

\begin{prop} For any $f \in \C$, define $Q^f: \B \rightarrow
\C$ by $Q^f_B = \chip{[f \in B]}$. Then $Q^f \in {\mathcal O}$. Conversely, if $Q \in
\mathcal{O}$ is any $\Q$-valued measure, then there exists a function $f \in \C$ such that
$Q = Q^f$. Thus, $ {\mathcal O} = \C$.   \qd
\end{prop}

This gives  a classical spectral theorem for $\C$ as follows:

\begin{prop} For any $f \in \C$, define $Q^f(\lambda) = \chi^{(X)}_{[f \leq \lambda]}.
\forall \lambda \in \R$.  Then one may write any $f \in \C$ in the following integral form:

\begin{equation}
f = \int_{-\infty}^{\infty} \lambda dQ^f(\lambda)
\end{equation}

\noindent  Furthermore,  for any continuous Borel function of $f$,

\begin{equation}
 g \circ f = \int_{-\infty}^{\infty} g(\lambda) dQ^f(\lambda).
\end{equation}

\end{prop}

\spg {\em Proof.} Eq.(4.1) follows from the fact that for all $x \in X,$ $Q^f(.)(x)$ is a
nondecreasing function on $\R$ \cite[Theorem III.8.7]{hewi65}). Eq.(4.2) is by Mackey's
axiom III.   \hspace{1cm} \qd

\spg Using the language from Hilbert spaces, we call the $\Q$-valued measure $Q^f$ the {\em
spectral measure} corresponding to the observable $f$, and eq. (4.1) the {\em spectral
decomposition} of $f$.

Birkhoff and von Neumann motivated their logic of quantum mechanics with the
epistemological judgment that ``Before a phase-space can become imbued with reality, its
elements and subsets must be correlated in some way with experimental propositions'' , {\em
i.e.,} with the Borel sets of the real line $\R$ and its products $\R^n$
\cite[p.825]{birk36}. The designation of  the space $X$ as the algebraic theory's ``phase
space'' is their terminology.  Each spectral measure $Q^f \in {\mathcal O}$ defines a
correlation of the Borel sets in $\R$ with sets in the algebraic phase space $X$ as
follows:

\begin{prop} For any $f \in \C$, the measure $Q^f \in {\mathcal O}$ is a
lattice homomorphism on ${\mathcal B}$ into the lattice $\Q$, transforming the operations
$(\subseteq, \bigcup,\bigcap, ^{\prime})$ to $(\leq, \bigvee, \bigwedge, ^{\prime})$ and
preserving set inclusion. Hence, for any $f \in \C$, the compose  $\phi \circ Q^f: \B
\rightarrow \B(X)$ is a lattice homomorphism on the Borel sets of $\R$ into $\Bo$, where
$\phi$ is the isomorphism defined in Proposition III.1.
\end{prop}

\noindent \prf Recall that $\Q$ is a complete lattice. One has $Q^f_{B_1 \bigcap B_2} =
\chip{[f \in B_1] \bigcap [f \in B_2]} = \chip{[f \in B_1]} \bigwedge$ $ \chip{[f \in
B_2]}$ and $Q^f_{B_1 \bigcup B_2} = \chip{[f \in B_1 \bigcup B_2]} = \chip{[f \in B_1]}
\bigvee \chip{[f \in B_2]}$ for the meet and join, and $Q^f_{B^{\prime}} = {\bf 1} -
\chip{[f \in B]}$ for complementation.  Furthermore, $E, F \in \B$, $E \subset F$, goes to
$\chip{E} \leq \chip{F}$. \hspace{5mm}\qd

\spg This establishes the role of the algebraic observables in the theory.

\vspace{3mm} \begin{center} {\bf B. States} \end{center}

The  probability of the set $[f \in B]$ in the initial state $\zemu \in \glosta$ is just
the expectation value with respect to the probability measure $\sigma_{\mu}$ of the
random variable $\chip{[f \in B]}$:

\begin{equation} \zemu(\chip{[f \in B]}) = \int_X \chip{[f \in B]}
d\sigma_{\mu} = \int_{[f \in B]}  d\sigma_{\mu} = \sigma_{\mu}([f \in B] )
\end{equation}

\spg This is the probability the measurement finds the system in a MC state $x$ belonging
to the set $[f \in B] \subseteq X$ when the  lattice is in state $\zemu$.  It is given in
terms of the spectral measures $Q^f$ by Mackey's Axiom I as follows:

\begin{prop} For any observable $f \in \C$ and any state
$\zemu \in \glosta$, the function $p$ defined by

\begin{equation} p(Q^f, \zemu,B) = \zemu(\chip{[f \in B]})   \end{equation}

\spg is a probability measure on $(\R, \B)$.   \qd
\end{prop}

\spg Following Haag and Kastler we call an algebraic theory a {\em complete} theory of
measurement if for all Borel sets $F \in \mathcal{B}(X)$, and for all algebraic states
$\zemu \in \glosta$ one can write the probability of finding the system in a MC state $x
\in F \subseteq X$, given that it is initially in the state $\zemu$, . To show that we
have a complete theory in this sense, set $f = \chip{F}$. Then $f \in \C$, and from eq.
(4.3), $\zemu(\chip{[f \in (1/2,3/2)]}) = \sigma_{\mu}(\{x \in F\})$. We have treated the
measurements themselves as represented by local observables $\wat{t}$ and their states
$E_t$ referred to a particular system $\Lambda_t$.   Haag and Kastler regard operations
of the form $f = \chip{F}$ as filters, passing the MC states in $F$ and blocking the
rest. Correspondingly, they call the probabilities $\sigma_{\mu}(\{x \in F\})$ {\em
transmission probabilities.}

\vspace{5mm} \noindent{\bf {\large V \hspace{4mm} Applications }} \setcounter{section}{5}
\setcounter{equation}{0} \setcounter{thm}{0}

\vspace{3mm} We conclude with two applications of the axiomatic theory. They pertain
especially to results on the equivalence of ensembles in the thermodynamic-limit program.
The first is a very basic question for the axiomatic theory itself: is the theory's set
of observables $\C$ large enough? Given any two states $\zemu, \zeta_{\nu} \in \glosta$,
is there an observable $f \in \C$ such that $\zemu (f) \neq \zeta_{\nu}(f)$? Since the
mapping in Proposition IV.3 is into, not onto, the correlation provided by a measure $Q^f
\in {\mathcal O}$ does not in general define a state $\zemu$ on all of $\B(X)$.
Nevertheless, it has the ability to distinguish two states $\zemu$ and $\zeta_{\nu}$ by
measurements, as follows.

\begin{prop} If $p(Q^f,\zemu,B)= p(Q^g, \zemu,B)$ for all $\mu \in
\threads$, $B \in \B$, then $f=g$.  Conversely, if $p(Q^f, \zemu,B) = p(Q^f, \zeta_{\nu},
B)$ for all $f \in \C$, $B \in \B$, then the states $\zemu = \zeta_{\nu}$.
\end{prop}

\spg {\em Proof.} Axiom III.  \qd

\spg That is, states separate observables, and observables separate states. $\C$ contains
points that do not represent measurements because $\W$ is the completion of $\winf$.
Nevertheless, the set W = $\psi_K \circ \Delta_K(\winf)$ is (strongly) dense in $\C$ (Part
A, Corollary II.12, Theorem II.15), so that if any $f \in C$ separates the states $\zemu,
\zeta_{\nu}$, we may construct a convergent sequence of functions $(g_n) \in$ W such that
$g_n \rightarrow f$, and it does contain points that separate these states.

It is noteworthy that this separation property does not conflict with results on the
equivalence of ensembles.  These theorems have to do with the convergence  of sequences of
local observables when the lattice is in one of the standard ensembles (MC, canonical,
grand canonical).  They show that the sequences converge in probability to the same limit
functions for all three ensembles \cite[Theorem A5.8]{lanf73}, {\em i.e.,} as the sizes of
systems get larger and larger, measurements of any quantity give the {\em same values} in
the three ensembles except possibly on  sets of configurations of decreasing probability.
But in general, $f^t \overset{P}{\rightarrow} f$ does not assure that $\int f^t d\mu_t
\rightarrow \int f d\mu$ unless there exists an integrable dominating function $g$ such
that $|f^t| \leq |g|$ for all $t \in \mathbf{J}$ (Lebesgue Dominated Convergence Theorem
\cite[Theorem 7.2.C]{loev63}). Thus, agreement of the limit functions does not assure
agreement of the limits of their expectation values. In physical terms, the dominating
function has the effect of excluding large fluctuations from the limiting value of an
observable.

For the second application, we show that there is a weak equivalence of states if one
allows some experimental error in measurements. Specifically, suppose the expectation value
of a particular observable $f \in \C$ in a given initial state $\zemu$ is only determined
(or estimated) to within an accuracy of $\zemu(f) \pm \varepsilon$. Then the measurement
cannot be used to separate $\zemu$ from any state $\zeta_{\nu}$ in the wk*-neighborhood of
$\zemu$ defined by the basic open set ${\mathcal N}(\zemu; f, \varepsilon) = \{\zeta_{\nu}:
|\zemu(f) - \zeta_{\nu}(f)| < \varepsilon\}$. With repeated measurements, one can estimate
the relative frequency (or probability) of a set $[f \in \B]$ to any degree of precision.
However, this cannot exclude these considerations with any finite number of measurements.
States close together in this sense are essentially physically equivalent.

\vspace{5mm} \begin{center} {\sc Acknowledgement}.
\end{center}

The author wishes to express his gratitude to Rudolf Haag for his many suggestions during
the writing of this manuscript.

\end{document}